# A Prior-Predictive Monte Carlo Framework for Pricing Complex Data Products in Data-Poor Markets

Adam L. Siemiatkowski[1], Victor Zhirnov[2,3], Kashyap Yellai[2], Gabriella Bein[2], Terresa Zimmerman[2]

[1]Columbia University, New York, NY, USA
[2]Semiconductor Research Corporation (SRC), Durham, NC, USA
[3]North Carolina State University, Raleigh, NC, USA

Pricing advanced data products - particularly in complex fields such as semiconductor manufacturing - is a fundamentally challenging task due to the sparsity of publicly available transaction data, and its frequent heterogeneity and confidentiality. While data value depends on multiple interacting factors, such as technical sophistication, quality, utility, and licensing rights, traditional pricing methods tend to rely on ad-hoc heuristics or require massive amounts of historical transaction data. In an increasingly data-based economy, we introduce a prior-predictive Monte Carlo framework that enables the generation of fair, consistent, and justified price ranges for data products in the absence of empirical data. By simulating many plausible pricing "worlds" and deal configurations, the framework produces stable probabilistic price bands (e.g., P5/P50/P95) rather than single point estimates, creating an auditable and repeatable probabilistic pricing system with business realism enforced via constraint-truncated priors. The proposed model bridges traditional data pricing rooted in professional experience with a data-based approach that also allows for classical Bayesian updating as more transaction data is accumulated.

## 1. Introduction

The rapid growth of data-driven decision making has turned data into one of the most important strategic assets an enterprise can use to gain advantage over its competitors (Brynjolfsson). That is perhaps most apparent in sectors of advanced technology, such as semiconductor manufacturing, where datasets related to inspection imagery, metrology signals, process logs, etc. can directly affect product quality and process efficiency (Brynjolfsson). Despite the significance of data transactions between companies, such datasets are nearly impossible to price as transactions are mostly confidential, highly customized (Koutris). On top of that, data products are not differentiated through simple factors such as size, instead being subject to variables such as technological sophistication, quality, and licensing rights - all carrying different weight depending on the use case of the buyer.

Currently, complex data products are frequently priced using arbitrary methods or informal multiplier tables (Koutris). While these approaches are certainly successful to a certain measure, as it is necessary for companies to establish the value of certain data-product attributes depending on their internal needs, they also suffer from issues such as producing inconsistent outcomes across similar transactions, provide no way of quantifying uncertainty and risk, and obscure the factors driving prices. More complex, regression or machine learning based models, offer no solution to these problems as their estimates rely on historical observations very rarely found in advanced technology markets and developing industries (Geman).

This paper, a development to the group's previous works, addresses that issue by postulating a prior-predictive deterministic Monte Carlo pricing framework designed explicitly for data-poor environments. Rather than inferring prices from past transactions, the proposed method formalizes experts' judgements about price drivers in semiconductor manufacturing data (technology node, process coverage, quality and freshness, utility, and licensing rights) into an auditable mathematical model. Price is expressed as a multiplicative function of these drivers, ensuring positivity, interpretability, and scale invariance. Uncertainty in baseline value, elasticities, and negotiation outcomes is represented probabilistically through a governed prior distribution, while business and fairness rules are enforced via explicit constraints. The use of the Monte Carlo architecture is then applied to determine the consequences of the assumptions across various "pricing worlds," yielding a probability distribution of prices for any given deal configuration (Metropolis).

## 2. Model Definition

We propose a prior-predictive, constraint-governed Monte Carlo pricing model that outputs probabilistic price bands for advanced data products (such as the discussed case study of semiconductor datasets) given:

1. An agnostic baseline price anchor $b_0$ derived from the previously discussed GDP estimate (Appendix A, Zhirnov)
2. Deterministic multiplier functions $x_j(\cdot)$ that encode economic factors and legal structures (technology node (TN), process coverage (COV), quality and freshness (QF), utility (UTIL), and rights (RIGHTS) in the discussed case study).
3. Uncertainty over elasticities $\beta_j$ and a residual noise factor $\sigma$
4. A probabilistic "mix of deals" over various product configurations
5. Additional constraints (later referred to as anchors) that enforce certain elements of business realism and monotonicity.

The output of the function is therefore not a single price, but a distribution of likely selling prices summarized by *P5*/*P50*/*P95* price bands.

## 3. Mathematical Architecture

1. *Notation and Objects*

Let each potential selling transaction $i \in \{1, ..., n\}$ be modelled by a vector of positive multipliers:

$$x_i = (x_{i,\,TN},\, x_{i,\,COV},\, x_{i,\,QF},\, x_{i,\,UTIL},\, x_{i,\,RIGHTS}) \in (0, \infty)^5$$

Let $p_i > 0$ signify the price per megabyte for deal $i$, and the parameter vector be defined as:

$$\theta = (\alpha, \beta, \sigma), \quad \beta = (\beta_{TN},\, \beta_{COV},\, \beta_{QF},\, \beta_{UTIL},\, \beta_{RIGHTS}), \quad \sigma > 0$$

with $\alpha \in \mathbb{R}$ and $\beta \in \mathbb{R}^5$.

The baseline is then anchored via:

$$\alpha = ln(b_0)$$

2. *Probability Space and Random Variables*

Let $(\omega, f, \mathbb{p})$ be a probability space on which all following random variables are defined on, with the parameter vector defined as:

$$\theta := (\alpha, \beta, \sigma) \in \Theta := \mathbb{R} \times \mathbb{R}^5 \times (0, \infty)$$

with an index set of pricing levers:

$$\mathbb{J} = \{TN, COV, QF, UTIL, RIGHTS\}$$

which when convenient can be identified with $\{1,\ldots,5\}$.

Let $A$ denote the aforementioned transaction data attributes with a formally defined attribute space $(A, \mathcal{G})$ where $A$ is the set of possible attribute tuples and $\mathcal{G}$ is an $\sigma$-algebra on $A$. In cases of mixed discrete and continuous attributes, $\mathcal{G}$ is the product $\sigma$-algebra.

The multiplier space is then defined as:

$$X := (0, \infty)^5$$

with a Borel $\sigma$-algebra $B(X)$.

### 3. Inputs Supplied in the Dataless Setting

i. A prior distribution over the parameters

A probability measure $\pi$ on $(\Theta, B(\Theta))$ must be provided. Writing $\theta \sim \pi$ then means that $\theta$ is an $\Theta$-valued random variable with law $\pi$. Typically:

$$\alpha \sim N(ln(b_0), s^2_\alpha)$$
$$\beta_j \sim N(\mu_j, s^2_j)$$
$$\sigma \sim \text{HalfNormal}(s_\sigma)$$

ii. A "mix of deals" distribution over attributes

A probability measure $P_A$ on $(A, \mathcal{G})$ must also be provided to describe how often each type of transaction appears in a given business pipeline. Let $A \sim P_A$.

iii. A deterministic and measurable transition from attributes to multipliers

Let $g$ be a function such that:

$$g : A \to X$$

where $g$ computes multipliers from attributes:

$$x := g(A) = (x_{TN}, x_{COV}, x_{QF}, x_{UTIL}, x_{RIGHTS})$$

It is crucial that $g$ is $\mathcal{G}/B(X)$-measurable so that $x$ is a random vector whenever $A$ is random. In practice, this is ensured by the fact that $g$ is almost always built from measurable primitives.

### 4. Economic Model

A practical structural economic model for price per megabyte is:

$$p_i = b_0 \prod_{j \in \mathbb{J}} x^{\beta_j}_{i,j} \varepsilon_i$$

This formulation is effective as it ensures the positivity of prices, multiplicative compounding of price drivers, and a direct interpretation of elasticity.

5. *Statistical Representation*

Taking a bijection on $(0, \infty)$ yields the following multiplicative model (Wooldridge):

$$ln(p_i) = \alpha + \sum_{j \in \mathbb{J}} \beta_j \, ln(x_{i,j}) + \eta_i, \quad \eta_i := ln(\varepsilon_i)$$

Assume that:

$$\eta_i \mid \theta \sim N(0, \sigma^2),$$

Equivalently (Limpert):

$$\varepsilon_i \sim \text{LogNormal}(0, \sigma^2).$$

Define:

$$Y := ln(P), \quad Z := ln(x)$$

Then the conditional model is:

$$Y \mid x, \theta \sim N(\alpha + \beta^\top Z, \sigma^2),$$

and,

$$P \mid x, \theta \sim \text{LogNormal}(\alpha + \beta^\top Z, \sigma^2).$$

This log-linear form is precisely equivalent to the multiplicative model, and is derived primarily for simulation efficiency.

6. *Constraints Defined as a Truncated Prior*

Business rules (monotonicity, premium caps, anchor inequalities) define a measurable subset of admissible parameters:

$$\Theta_C \subseteq \Theta$$

The governed prior is the conditional distribution (Bernardo):

$$\pi_C(\cdot) := \pi(\cdot \mid \theta \in \Theta_C),$$

Sampling from $\pi_C$ via rejection sampling is exact. All accepted draws are i.i.d. from $\pi_C$. This is proven by:

$$P(\theta_{acc} \in B) = P(\theta_{acc} \in B \mid \theta \in \Theta_C) = \pi(B \cap \Theta_C)/\pi(\Theta_C) = \pi_C(B)$$

## 7. Prior-predictive Distribution of Prices

The unconditional distribution of prices is the prior predictive law:

$$\mathcal{L}(P) = \int_\Theta \int_{CA} \mathcal{L}(P \mid x = g(a), \theta) \, \mathsf{P}_A(da) \, \pi_C(d\theta)$$

The above is a mixture of lognormal distributions mixed over the uncertainty in parameters $\theta$ and the uncertainty in transaction configurations $A$.

For any configuration class $C \subseteq A$ (e.g., "5 nm, partial exclusivity, 24 months"), the conditional price law is:

$$\mathcal{L}(P \mid A \in C) = \int_\Theta \int_{CC} \mathcal{L}((P \mid x = g(a), \theta) \, \mathsf{P}_A(da \mid A \in C) \, \pi_C(d\theta)$$

## 8. Sampling Scheme

First, a triangular array of random variables is taken:

A parameter is drawn such that:
$$\{P^{(t,i)} : t = 1, \ldots, T; \ i = 1, \ldots, N\}$$

$$\theta^{(t)} = (\alpha^{(t)}, \beta^{(t)}, \sigma^{(t)}) \sim \pi_C,$$

independently for each $t$.

For each $t$, attributes are drawn:

$$A^{(t,i)} \sim \mathsf{P}_A(\cdot \mid A \in C), \quad i = 1, \ldots, N,$$

independently across $i$.

The multipliers are then computed as:

$$x^{(t,i)} = g(A^{(t,i)}).$$

And noise is sampled

$$\eta^{(t,i)} \mid \theta^{(t)} \sim \mathcal{N}\big(0, (\sigma^{(t)})^2\big)$$

independently across $i$.

Finally, the price can be defined as:

$$P^{(t,i)} = \exp\!\big(\alpha^{(t)} + (\beta^{(t)})^\top \ln x^{(t,i)} + \eta^{(t,i)}\big)$$

It is important to note that each P$^{(t,\,i)}$ has the same distribution as P $\mid A \in C$, and the family $\{P^{(t,\,i)}\}$ is i.i.d. because each draw uses the same prior $\pi_C$ and conditional attribute law (Bahadur). This guarantees convergence.

9. *Monte Carlo Estimator*

Let $h : (0, \infty) \to \mathbb{R}$ be a measurable function such that:

$$\mathbb{E}\big[\,|h(P)|\,\big|\,A \in C\big] < \infty$$

Then the Monte Carlo estimator takes the form of (Metropolis):

$$\widehat{\mathbb{E}}_{T,N}[h(P)] := \frac{1}{TN} \sum_{t=1}^{T} \sum_{i=1}^{N} h\left(P^{(t,i)}\right)$$

Because P$^{(t,\,i)}$ are i.i.d. with finite expectations, the Strong Law of Large Numbers suggests that (Billingsley):

$$\widehat{\mathbb{E}}_{T,N}[h(P)] \xrightarrow{\text{a.s.}} \mathbb{E}[h(P) \mid A \in C] \quad \text{as } TN \to \infty.$$

Therefore, it is certain that Monte Carlo averages converge to the true prior-predictive expectation.

In reality, however, prices are more dependent on quantiles, not expectations themselves. Therefore, let $F$ denote the cumulative distribution function of P $\mid A \in C$, and define the quantile:

$$Q(q) := \inf\{p : F(p) \geq q\}$$

Then let $\widehat{Q}_{T,N}(q)$ be the empirical $q$-quantile of the sample $\{P^{(t,\,i)}\}$. It then follows from the standard theory of empirical quantiles that if $F$ is continuous and monotonously increasing (Bahadur):

$$\widehat{Q}_{T,N}(q) \xrightarrow{\text{a.s.}} Q(q) \quad \text{as } TN \to \infty$$

Thus, the Monte Carlo estimator leads to consistent estimators of governed pricing bands.

## 4. Hypothetical Semiconductor Dataset Case Study

1. *Outline of the Hypothetical Scenario*

Suppose an advanced foundry enterprise is conducting a transaction with a fabless chip enterprise concerning the manufacturing and analysis of 3 nm chips. The dataset in question compiles information about a number of fabrication processes, namely inline metrology time-series, defect inspection imagery, parametric tests (WAT/PCM) + wafer-sort yield outcomes, tool context metadata (tool IDs, chambers, recipe version), as well as labels: "good/bad" yield outcomes; defect classes; process drift markers.

The transaction also includes the following delivery and rights frameworks: 5 PB of compressed imagery and structured logs (≈ 5,120 TB ≈ 5,242,880 GB ≈ 5,368,709,120 MB) delivered as an encrypted cloud bucket, weekly refresh for 3 months, then static snapshot. The datasets are non-exclusive, however, can be used for enterprise-wide use, internal derivative models (they can train internal ML models) with no redistribution/sub-licensing. The term period is 24 months. The proposed method requires the following inputs from a user: a baseline price anchor (Appendix A, Zhirnov), a technology node multiplier, a process coverage multiplier, a coverage and freshness multiplier, an utility multiplier, and licensing rights multiplier.

It is now crucial to outline the rules of how the multipliers should be selected in practice. An enterprise should select the multipliers in a semi-arbitrary way, meaning that the values do not have an inherently defined meaning, however, they should be a reflection of the buyer's perceived value of an incremental improvement in the attribute in question, with the base product always having a multiplier of 1. For instance, in the hypothetical case study described below, the technology node multiplier for a 3 nm chip fabrication process is 1.65, however, say that the buyer's data needs are more or less ignorant to the technology node of the fabricated chips, that multiplier should be significantly lower. On the other hand, if the buyer's priority is to train ML models on the purchased data, the multiplier on derivative rights should likely be higher than in the case study where it is set as 1.3.

### 2. Multiplier Magnitude Selection

a) A baseline Price Anchor

$b_0$ has been already selected as a value that all data holds (see Appendix A for a full thought process), agnostic of its properties, which has been discussed in this group's prior publication (Zhirnov). For 2026, the figure is $3.90 \times 10^{-5}$ USD/MB, hence that will be treated as the baseline price in this case study as well.

b) Technology Node Multiplier

This is an enterprise-specific feature of the data. A user of the model should select arbitrary figures that roughly relate the technological advancement of the processes the data is concerned with according to their means of production and business model. For the sake of this case study, the following multipliers are chosen:

$$10 \text{ nm} \rightarrow 1.25$$
$$7 \text{ nm} \rightarrow 1.35$$
$$5 \text{ nm} \rightarrow 1.50$$
$$3 \text{ nm} \rightarrow 1.65$$
$$2 \text{ nm} \rightarrow 1.80$$

Hence, the TN multiplier for the case study is equal to 1.65

c) Coverage Multiplier

As is the case with all multipliers, their specific values should be established based on experts' opinions, however, a reasonable formula for the magnitude of a coverage multiplier could be:

$$x_{COV} = 1 + 0.15\ln(1+m)$$

where $m$ is the number of covered processes of interest for the buyer. The formula takes a logarithmic form because the number of covered processes is usually associated with a diminishing (Varian) return best modelled by a logarithmic function. In this case, $m = 6$, hence the multiplier is 1.2919.

d) Quality and Freshness Multiplier

A reasonable formula determining the quality and freshness of a product would be:

$$x_{QF} = 0.85 + 0.2q + 0.1c + 0.1(1 - age/24)$$

where $q$ is a quality score from 0 to 1 and $c$ is the completeness score from 0 to 1. The terms accompanying $c$ and $q$ should be selected specifically to the transaction, however, in our case study it can be assumed that the quality of the data is significantly more important than its completeness. Being that the data is 6 months old, the quality and freshness multiplier is 1.21.

e) Utility Multiplier

Utility is best described as the profit the buyer can gain from the purchase of the data. In the case of even a small yield improvement (+0.5%–2%) on high-volume production can translate into tens of millions in annual savings or incremental revenue. A high impact sale discussed in our case study could plausibly yield a benefit of 25 million USD,

$$x_{UTIL} = 1 + 0.4\log(1+V/10^6)$$

where $V$ is the estimated profit coming from the data purchase. With an assumed $V = 25 \times 10^6$, the utility multiplier becomes 1.566.

f) Rights Multiplier

The rights structure of the deal is layered as it is:

- non-exclusive = 1.0: baseline; buyer has access but competitors can also license.
- derivatives = 1.3: allowing internal model training increases ROI significantly
- 24 months = 1.1: longer term means longer value capture
- enterprise = 1.15: enterprise use increases the organization-wide footprint

Hence, the total rights multiplier amounts to:

$$x_{RIGHTS} = (1.0)(1.3)(1.1)(1.15) = 1.6445$$

3. *Calculating Prices*

The postulated method defines a deterministic mapping:

$$x = g(A) = (x_{TN}, x_{COV}, x_{QF}, x_{UTIL}, x_{RIGHTS}) = (1.65, 1.29189, 1.21, 1.56599, 1.6445)$$

Parameter uncertainties defined in Section 3 are the following for the case study assuming conservative uncertainties for the baseline price (0.25 in log space) and variance caused by factors not included in any of the multipliers (0.35 in log space):

$$\alpha \sim \mathcal{N}(\ln b_0,\ 0.25^2)$$
$$\sigma \sim \text{HalfNormal}(0.35)$$

The $\beta$ values in this case study should reflect the belief that utility and rights often drive price more than coverage, coverage has diminishing returns (hence < 1), and that quality is near proportional ($\approx$ 1). Example values embodying this perspective are:

$$\mu = (1.17, 0.86, 0.97, 1.39, 1.16)$$

With standard deviations:

$$s = (0.15, 0.12, 0.10, 0.18, 0.14)$$

As mentioned in Section 3, business rules define $\Theta C \subseteq \Theta$ and we sample from $\pi_C(\cdot) := \pi(\cdot \mid \theta \in \Theta_C)$ via rejection sampling. For the case study, we enforce:

- $0 < \beta_j < 3$, to ensure that there are no negative elasticities or unrealistic blowups
- $0 < \sigma < 1$

Then, a $T$ (number of parameter draws ("worlds")) and $N$ (number of noise draws per world) must be selected for the Monte Carlo simulation. For the purpose of the case study, let $T = 5000$ and $N = 10$.

Now, with the price formula derived in Section 3:

$$ln(p_i) = \alpha + \sum_{j \in \mathbb{J}} \beta_j \, ln(x_{i,j}) + \eta_i, \quad \eta_i := ln(\varepsilon_i),$$
$$p = \exp(ln(p)),$$

and the previously computed:

$$x = (1.65, 1.29189, 1.21, 1.56599, 1.6445),$$

giving:

$$ln(x) \approx (0.5008, 0.256, 0.1906, 0.447, 0.497)$$

Conducting a point estimate using the mean elasticities (through setting $\varepsilon = 1$, and $\beta = \mu$) gives:

$$ln(p_{point}) = ln(b_0) + \sum_j \mu_j \, ln(x_j)$$

Finally resulting in:

$$ln(p_{point}) = -10.15195 + 2.192 = -7.960$$
$$p_{point} = \exp(-7.960) = 3.49 \times 10^{-4} \text{ USD/MB}$$

This procedure is then repeated many times, drawing $\alpha, \beta, \sigma$ from the prior (rejected if constraints fail), as well as $\eta_i = N(0, \sigma^2)$. For each draw, $ln(p)$ and $\exp(ln(p))$ are computed, all $p$ values are collected and reported as a distribution.

4. *Case Study Results*

Running the simulation on the specific set of parameters described in this case study results in the following output:

**P5:** $1.6926 \times 10^{-4}$ USD/MB

**P50:** $3.4630 \times 10^{-4}$ USD/MB

**P95:** $7.0578 \times 10^{-4}$ USD/MB

Translating to GB:

**P5:** $0.173 / GB

**P50:** $0.355 / GB

**P95:** $0.723 / GB

Giving a total contract value of:

**P5 total:** $0.909M

**P50 total:** $1.859M

**P95 total:** $3.789M

If instead of pricing one deal, one would want to model an entire sales pipeline, you would additionally specify $P_A$ (how often different deal types occur) and then you would sample $A \sim P_A$ before computing $x = g(A)$.

## 5. Discussion

1. *Efficiency of the Architecture and Comparison to Alternatives*

The most common alternative to the applied Monte Carlo method is creating a neural network or iterative machine learning model. Further, it is obvious that these models could be more effective at price estimation, however, the context of the work makes a log-linear architecture more beneficial for numerous reasons.

Once $\theta$ is drawn, pricing a deal is $O(|\mathbb{J}|)$ arithmetic operations, so for $N$ deals and $T$ parameter worlds, the runtime of the simulation is $O(TN|\mathbb{J}|)$, which is linear in the number of simulations. On the other hand, a machine learning model would require iterative optimization and multiple epochs, which even after ignoring constant factors, is significantly heavier in evaluation. Furthermore, after real transaction data is collected and an estimate is made by Ordinary Least Squares, the problem remains convex with a unique minimizer when $X^TX$ is invertible, whereas nonconvex models are likely to have several minima. The application of the postulated model is a fair pricing system between sellers and buyers, eventually aiming at maximizing profits for data transactions. Any machine learning model would implicitly be a black box that doesn't inform the parties involved in the deal with the actual justification for a proposed selling price.

Machine learning models (e.g. Random Forests, Neural Networks) are also only efficient in data-rich environments (Still), which is a rare situation in the case of young and quickly developing markets. In a data-poor situation, machine learning models are not identifiable because they have too many degrees of freedom (Still). Additionally, they do not enforce monotonicity unless explicitly constrained, and in the case of deep learning networks this is a nontrivial problem (Still). The proposed model behaves well in dataless or data-poor environments, as it is initially informed by expert opinions, and has a significantly lower variance. Further, if the model is fitted with OLS on log data under Gauss-Markov assumptions, $\beta$ becomes the best linear unbiased estimator of the elasticities, i.e. it minimizes variance among all linear unbiased estimators (Aitken).

Another alternative would be a simpler solution for modelling the price such as:

$$p = cost + margin$$

This model, however, fails to consider any extreme cases, and is therefore bound to occasionally underprice a valuable asset or overprice a generic one.

### 2. *Possibilities of Further Developing the Model*

A big advantage of the model is that with the accumulation of more transaction data by an enterprise, it can be transformed into a posterior-predictive model without many structural changes via Bayesian updating (Gelman, Murphy). The prior distribution $\pi(\theta)$ could then be substituted be observed deals, which would cause the posterior distribution over $\alpha$, $\beta$ to contract around the empirically determined values. This would also allow to more accurately determine the distributions of the multiplier magnitudes as some of them could have been previously over or under weighted.

Another exciting modification could be the decomposition of the $\sigma$ factor, which as it stands models all undefined variability, however, as the history of transaction increases, more components could be added such as urgency effect, or contract size effect (Limpert). Unexplained noise is therefore possible to explain by additional multipliers. If that was the case, a reasonable question would be: why not transition to regression or machine learning models? While this may seem as the better solution, even with large amounts of data, naive regression and ML models suffer from regions of data sparsity, meaning that some deal configurations remain so rare that they can be mistakenly treated as similar to a previously seen transaction example, whereas in reality they are vastly different. In probabilistic models like the one proposed in the paper, however, with an increased data size a reasonable solution would be to utilize the technique of partial pooling across various transaction cases. As the model is transitioning in hierarchical Bayesian prices, information can be shared between similar deal cases, which gives better estimations in the situation of some cases remaining extremely rare.

Crucially, the model could be extrapolated to almost any data product - not just semiconductor manufacturing. If there is a sound way of assigning price driving attributes for it, the model is applicable to any dataset.

### 3. *Model Limitations*

Without changing the fundamental architecture of the model, it is important to state its inherent limitations. The model will never be able to be used to discover "true" market prices as it fully depends on the initial input of expert judgements. This is not necessarily a shortcoming as because of that feature it is better suited for discovering a reasonable selling price, however, that is a very different metric than an objective market value. Connected to the fact that the model is based on expert judgements is perhaps its biggest risk - poor priors will result in poor results. If the model's outputs are absurd or not satisfactory, it is likely due to the fact that the inputs were poorly constructed, however, that issue may be difficult to solve. In short, the model's performance is still highly dependent on human judgement, and this judgement is not eliminated at any point; it is only mathematically formalized.

# 6. Appendix

**APPENDIX A: Global Data Economy (2015–2035)**

The purpose of this analysis is to establish a segmented framework as a coherent economic boundary for the Global Data Economy, enabling consistent historical reconstruction and forward projection against Global GDP through 2035. This approach allows us to estimate a lower boundary on the average price for any data agnostic of its content and properties (in this paper, it is referred to as the agnostic baseline price anchor $b_0$).

This analysis defines and quantifies the Global Data Economy as a distinct, measurable component of Global GDP. The objective is to establish a consistent economic framework for sizing, segmenting, and tracking data-driven industries across time, enabling comparison of digital value creation to total world economic output.

To achieve this, we:
- Identified the principal markets in which data is produced, stored, processed, distributed, and monetized.
- Reduced overlap and double-counting by assigning each activity to a single primary economic function.
- Selected ten core industry segments that together represent the full data value chain (Table 1).
- Used market size (revenue) as the standard metric for all segments, with the exception of Cryptocurrency, where market capitalization is used to reflect its role as a monetary and store-of-value system.

Where available, historical market data was used; where gaps existed, values were reconstructed using published CAGR ranges. Forward projections were modeled through 2035 to align with long-range CAGR and GDP outlooks. Earlier periods (pre-2020) required normalization due to inconsistent segment definitions, emerging market status, and aggregation of digital revenues into broader ICT and industry categories (Table 2).

The resulting model enables analysis across four integrated dimensions:
1. Global GDP
2. Total Data Economy Value (USD)
3. Global Data Volume (Zettabytes)
4. Economic Efficiency Metrics (Data per Dollar, Cost per MB, and % of GDP) (Table 3)

*1. Data Economy Market Segments*

The Global Data Economy is structured into ten segments spanning content, platforms, software, infrastructure, analytics, and digitally native financial and geospatial systems:

1. Cryptocurrency – Decentralized digital monetary and transactional infrastructure

2. Digital Media – Commercial creation and distribution of digital audio-visual and interactive content

3. Social Media Platforms – User-generated content networks and digital advertising ecosystems

4. Software – Application, platform, and infrastructure software enabling data creation and processing

5. Data Storage – Physical and cloud infrastructure for data persistence and access

6. Data Analytics – Platforms and tools transforming raw data into decision intelligence

7. Digital Health – Data-driven healthcare delivery, monitoring, and clinical intelligence systems

8. Digital Newspapers & Magazines (e-Press) – Digitized journalism and periodical publishing

9. eBooks – Digital long-form text publishing and distribution

10. Geospatial Imaging – Earth-observation, mapping, navigation, and spatial intelligence systems

Together these segments represent the core economic mechanisms by which data is generated, processed, secured, analyzed, distributed, and monetized at global scale, forming the measurable backbone of the modern digital or data economy.

## 2. Market Segment Profiles

### 1. Cryptocurrency

Segment Description:
Cryptocurrency represents the monetary layer of the data economy, enabling decentralized value storage, transfer, and settlement using cryptographic and distributed-ledger technologies.

Definition:
This segment includes blockchain-based currencies, transaction networks, mining and validation infrastructure, wallets, exchanges, and payment systems. It excludes traditional electronic banking and centralized payment rails.

Core Source:
CoinMarketCap – Cryptocurrency Prices, Charts, and Market Capitalizations

Key Players / Networks:
Bitcoin, Ethereum, Binance, Ripple, Tether, Solana, Cardano, TRON, Avalanche, Coinbase, Bitmain, NVIDIA, AMD

### 2. Digital Media

Segment Description:
Digital Media captures the commercial production, distribution, and monetization of video, audio, text, and interactive content across streaming, advertising, and on-demand platforms.

Definition:
Includes OTT streaming, digital advertising media, online video, audio streaming, interactive content, and enterprise digital publishing platforms. Excludes user-generated social networking (covered in Social Media Platforms).

Core Source:
Grand View Research – Global Digital Media Market

Key Players:
Apple, Amazon, Netflix, Disney, Sony, Paramount, Fox, AT&T, Thomson Reuters, Kaltura

### 3. Social Media Platforms

Segment Description:
Social media platforms are the primary global systems for user-generated data creation, social interaction, and targeted digital advertising.

Definition:
Online networks enabling profile creation, content sharing, messaging, video distribution, and community interaction, monetized primarily through advertising and data services.

Core Source:
The Business Research Company – Social Media Platforms Global Market Report

Key Players:
Meta (Facebook, Instagram), YouTube, TikTok, LinkedIn, X (Twitter), Snapchat, Reddit, Pinterest, Tencent, Douyin

### 4. Software

Segment Description:
Software is the computational engine of the data economy, providing operating systems, applications, and platforms that generate, manage, and process digital information.

Definition:
Includes enterprise, consumer, cloud, security, analytics, and infrastructure software across SaaS, PaaS, and on-premise deployments.

Core Source:
Grand View Research – Global Software Market

Key Players:
Microsoft, SAP, Oracle, IBM, Adobe, VMware, Salesforce, Intuit, ServiceNow, Alphabet

### 5. Data Storage

Segment Description:
Data Storage constitutes the physical and virtual infrastructure layer that preserves global digital information and enables high-availability access.

Definition:
Covers cloud storage, hyperscale data centers, enterprise storage systems, backup, archiving, and software-defined storage platforms.

Core Source:
Fortune Business Insights – Global Data Storage Market

Key Players:
AWS, Microsoft Azure, Google Cloud, Dell, HPE, NetApp, Pure Storage, Seagate, Western Digital, Equinix

### 6. Data Analytics

Segment Description:
Data Analytics transforms raw data into economic value through insight generation, prediction, and automation.

Definition:
Includes descriptive, predictive, prescriptive, and real-time analytics platforms, AI/ML analytics, and business intelligence systems.

Core Source:
Grand View Research – Global Data Analytics Market

Key Players:
IBM, SAP, Oracle, Salesforce, Amazon, Google, ThoughtSpot, Sisense, Mu Sigma, Zoho

### 7. Digital Health

Segment Description:
Digital Health applies data, connectivity, and analytics to clinical care, diagnostics, monitoring, and life-science operations.

Definition:
Includes telehealth, remote patient monitoring, digital therapeutics, AI diagnostics, interoperable EHRs, and health cloud platforms.

Core Source:
MarketsandMarkets – Global Digital Health Market

Key Players:
Philips, Medtronic, Abbott, Dexcom, Apple, AWS, Microsoft, IBM, Mayo Clinic, Cleveland Clinic

## 8. Digital Newspapers & Magazines (e-Press)

Segment Description:
e-Press represents the digitization of news and periodical publishing, subscription platforms, and advertising-supported digital journalism.

Definition:
Includes online newspapers, digital magazines, and associated advertising and subscription platforms.

Core Source:
Grand View Research – Digital Newspapers & Magazines Market

Key Players:
The New York Times, News Corp, Axel Springer, Schibsted, Daily Mail Group, Zinio, Readly, Magzter

## 9. eBooks

Segment Description:
eBooks are the digital distribution channel for long-form text content across consumer, professional, and educational markets.

Definition:
Includes consumer, educational, and professional eBook publishing and distribution platforms.

Core Source:
Market.us – Global eBooks Market

Key Players:
Amazon Kindle, Apple Books, Kobo, Barnes & Noble, Scribd, Lulu, Smashwords, Macmillan

## 10. Geospatial Imaging

Segment Description:
Geospatial Imaging provides the spatial data layer of the digital economy, supporting navigation, infrastructure planning, defense, climate monitoring, and autonomous systems.

Definition:
Includes satellite, aerial, drone, LiDAR, and mobile mapping platforms and spatial analytics software.

Core Source:
Fortune Business Insights – Global Geospatial Imaging Market

Key Players:
ESRI, Hexagon, Trimble, Fugro, Cyient, NV5, Woolpert, PASCO, Maxar, Planet Labs

**Table 1: Data Economy Market Segments**

| Segment | Metrics (cap/size) | $B (2025) | Reference |
|---|---|---|---|
| Cryptocurrency | cap | 3110.0 | https://coinmarketcap.com/ |
| Digital Media | size | 1147.0 | https://www.grandviewresearch.com/industry-analysis/digital-media-market-report |
| Social Networks | size | 1000.2 | https://www.thebusinessresearchcompany.com |
| Software | size | 838.5 | https://www.grandviewresearch.com/press-release/global-software-market |
| Data Storage | size | 255.3 | https://www.fortunebusinessinsights.com/data-storage-market-102991 |
| Digital Health | size | 199.1 | https://www.marketsandmarkets.com/Market-Reports/digital-health |
| Data Analytics | size | 94.7 | https://www.grandviewresearch.com/horizon/outlook/data-analytics-market-size |
| e-Press | size | 40.8 | https://www.grandviewresearch.com/industry-analysis/digital-newspapers-magazines-market |
| eBooks | size | 20.5 | https://market.us/report/ebooks-market/ |
| Geospatial Imaging | size | 2.8 | https://www.fortunebusinessinsights.com/geospatial-imaging-market-107834 |
| **Total** | | **6708.9** | |

**Table 2: Global Data Economy Market in 2015-2025**

| Segment | $B (2025) | $B (2024) | $B (202) | $B (2022) | $B (2021) | $B (2020) | $B (2019) | $B (2018) | $B (2017) | $B (2016) | $B (2015) |
|---|---|---|---|---|---|---|---|---|---|---|---|
| Cryptocurrency | 3110.0 | 3270.0 | 1640.0 | 799.0 | 2200.0 | 741.0 | 187.0 | 123.0 | 538.0 | 17.0 | 7.0 |
| Digital Media | 1147.0 | 939.6 | 833.0 | 740.0 | 650.0 | 580.0 | 520.0 | 460.0 | 410.0 | 360.0 | 320.0 |
| Social Networks | 1000.2 | 745.0 | 553.0 | 411.0 | 305.0 | 227.0 | 168.0 | 125.0 | 93.0 | 69.0 | 51.0 |
| Software | 838.5 | 730.0 | 600.0 | 517.0 | 860.0 | 810.0 | 740.0 | 740.0 | 680.0 | 620.0 | 560.0 |
| Data Storage | 255.3 | 218.0 | 230.0 | 217.0 | 170.0 | 150.0 | 135.0 | 125.0 | 120.0 | 110.0 | 100.0 |
| Digital Health | 199.1 | 162.1 | 180.0 | 143.0 | 268.0 | 175.0 | 140.0 | 112.0 | 89.6 | 71.7 | 57.3 |
| Data Analytics | 94.7 | 69.5 | 57.8 | 49.0 | 42.6 | 38.3 | 35.1 | 32.9 | 26.7 | 21.8 | 17.7 |
| e-Press | 40.8 | 38.7 | 37.2 | 35.7 | 34.2 | 32.8 | 31.4 | 30.1 | 28.8 | 27.6 | 26.5 |
| eBooks | 20.5 | 19.1 | 18.2 | 13.2 | 14.0 | 12.8 | 12.7 | 12.1 | 11.5 | 10.9 | 10.4 |
| Geospatial Imaging | 2.8 | 2.4 | 2.0 | 1.7 | 2.0 | 1.9 | 1.7 | 1.6 | 1.4 | 1.3 | 1.1 |
| **Total** | **6709.0** | **6194.5** | **4151.2** | **2926.6** | **4545.8** | **2768.8** | **1971.0** | **1761.6** | **1999.1** | **1309.2** | **1151.0** |

Table 3: Global Data Economy Market vs. Global GDP

| Year | Data Economy ($T) | GDP($T)[*] | % of GDP | ZByte[**] | $b_0$ ($/Mbyte) |
|---|---|---|---|---|---|
| 2015 | 1.2 | 75.8 | 1.5% | 12 | 9.59E-05 |
| 2016 | 1.3 | 77.1 | 1.7% | 19 | 6.96E-05 |
| 2017 | 2.0 | 82 | 2.4% | 24 | 8.27E-05 |
| 2018 | 1.8 | 87 | 2.0% | 33 | 5.34E-05 |
| 2019 | 2.0 | 88.3 | 2.2% | 24 | 8.15E-05 |
| 2020 | 2.8 | 86.1 | 3.2% | 59 | 4.69E-05 |
| 2021 | 4.6 | 98.2 | 4.6% | 24 | 1.88E-04 |
| 2022 | 3.0 | 102.4 | 2.9% | 100 | 2.93E-05 |
| 2023 | 4.2 | 106.9 | 3.9% | 120 | 3.46E-05 |
| 2024 | 6.2 | 111.1 | 5.6% | 147 | 4.21E-05 |
| 2025 | 6.7 | 117.2 | 5.7% | 175 | 3.83E-05 |
| 2026 | 7453.0[**] | 123.6[**] | 6.0% | 231 | 3.22E-05 |
| 2027 | 8864.1[**] | 129.6[**] | 6.8% | 297 | 2.98E-05 |
| 2028 | 10542.5[**] | 136[**] | 7.8% | 394 | 2.68E-05 |
| 2029 | 12538.6[**] | 142.6 | 8.8% | 491 | 2.55E-05 |
| 2030 | 14912.7[**] | 149.6 | 10.0% | 612 | 2.44E-05 |
| 2031 | 17736.3[**] | 155.5[**] | 11.4% | 811 | 2.19E-05 |
| 2032 | 21094.5[**] | 162.9[**] | 12.9% | 1042 | 2.02E-05 |
| 2033 | 25088.6[**] | 170.7[**] | 14.7% | 1339 | 1.87E-05 |
| 2034 | 29838.9[**] | 178.8[**] | 16.7% | 1721 | 1.73E-05 |
| 2035 | 35488.7[**] | 187.4[**] | 18.9% | 2142 | 1.66E-05 |

[*]Global GDP 2025| Statista

[**]Global Data_Statista

[***]Projection

**APPENDIX B: Monte Carlo Engine**

https://www.online-python.com/arRMbkCSnB